\documentclass[11pt,aps,prl,superscriptaddress,byrevtex]{revtex4}
\usepackage{times}
\begin{document}

\title{Comment on ``Some non-conventional ideas about algorithmic complexity''}

\author{David Poulin}
\email{dpoulin@iqc.ca}
\address{School of Physical Sciences, The University of Queensland,  Queensland 4072, Australia }

\author{Hugo Touchette}
\email{htouchet@alum.mit.edu}
\address{School of Mathematical Sciences, Queen Mary, University of London, London E1 4NS, UK}
\date{\today}
\begin{abstract}
We comment on a recent paper by D'Abramo [Chaos, Solitons \&
Fractals, 25 (2005) 29], focusing on the author's statement that an algorithm
can produce a list of strings containing at least one string whose
algorithmic complexity is greater than that of the entire
list. We show that this statement, although perplexing, is not as
paradoxical as it seems when the definition of algorithmic complexity is
applied correctly.
\end{abstract}
\maketitle

D'Abramo has advanced in a recent paper \cite{abramo2005} a number of ideas
related to the notion of algorithmic complexity (also called
Kolmogorov or Chaitin complexity \cite{li1997}), and what seems
to be a paradox related to the definition of this quantity. In
our opinion, D'Abramo's paper may leave the impression
that the very definition of algorithmic complexity is
contradictory or inconsistent as a result of this apparent paradox.
This, in fact, is not the case, as we would like to show now.

To start, let us recall the central point of D'Abramo's considerations
\cite[p.~29]{abramo2005}. Consider a program $p$ that outputs all binary
strings of length less than or equal to $N$ bits in lexicographic order,
that is, that outputs the sequence
\begin{widetext}
\[
\ell=0\ \ 1\ \ 00\ \ 01\ \ 10\ \ 11\ \ 000\ \ 001\ \ 010\ \ 100\ \ 110\ \ 101\
\ 011\ \ 111\ldots \underbrace{11\cdots 1}_{N {\rm\ bits}}.
\]
\end{widetext}
The program $p$ must obviously contain the number $N$
(to know when to halt), plus some other bits of instructions for
generating the strings and printing them. As a result, $p$ is of
length $\lceil \log _2N\rceil +k$ bits when written in binary
notation, where $k$ is some $N$-independent constant that accounts for the
overhead instructions. From this reasoning,
we conclude that the algorithmic complexity of the list $\ell$ is
at most $\lceil \log _2N\rceil +k$ bits. (Note that the number
$N$ could be describable with less than $\log_2 N$ bits.)

Now comes the apparent paradox. The number of binary strings contained in
$\ell$ is $2^{N+1}-2$, while the total number of programs having length at most $\lceil
\log _2N\rceil +k$ is $2^{\lceil \log _2N\rceil +k+1}-2$. For $N$ sufficiently large,
we have
\[
2^{N+1}-2\gg 2^{\lceil \log _2N\rceil +k+1}-2,
\]
which means that $\ell$ contains a lot more binary strings than what can be
produced by all programs of length at most $\lceil \log _2N\rceil +k$. Consequently,
there must be a least one string $s$ in $\ell$ having a complexity greater than
$\lceil \log _2N\rceil +k$, since there must be at least one string in $\ell$
which is not produced by a program of length at most $\lceil \log _2N\rceil +k$.
This is the contradiction or ``paradox'' that D'Abramo alludes to:\ $\ell$ contains
a string $s$ of complexity greater than the complexity of $\ell$ itself, that is,
greater than the length of the minimal program that generates it. In
D'Abramo's words \cite[p. 30]{abramo2005}: ``Suddenly a paradox appears: an
algorithm [...] is able to write a list of strings which contains at least
one that is more complex than [...] the algorithm itself.''

This last statement is quite perplexing. If a program could indeed output
something more complex that itself, then the very
definition of algorithmic complexity would be
inconsistent. Fortunately, this is not the case, and there is a precision to
be made here. The fact is that D'Abramo's program does {\em not} output
the string $s$ that has a complexity greater than that of $\ell$---it outputs
much more. Therefore, it is misleading to allude to the complexity of $s$ in
the context of the program $p$, and then compare it with the
complexity of $\ell$; $p$ generates $\ell$ not $s$.

To define the complexity of $s$, we need to find a minimal program that
produces it. Following D'Abramo, one can attempt to modify
the program $p$ so that it first produces $\ell$, and then selects within it a
given string $s$ as the final output \cite[p. 31]{abramo2005}: ``[the program] can then stop the
counter whenever it wants and print the last enumerated number. The counter could
be provided with a sort of counting completeness indicator, which would give the percentage
of the whole count reached till that moment.''
In order to work, this new program needs to contain $N$ as before, but also requires the
address of the string $s$,  which in this case corresponds to the string $s$ itself.
In other words, to tell the ``counter'' when to stop and print the final output
requires a number of bits equal to the shortest description of the desired output $s$.
Hence, the new program is at least as long as the shortest description of $s$,
and so what appeared at first to be a paradox now has a simple explanation.

The same conclusion can be reached from a different perspective by noting the
nonadditivity of the algorithmic complexity, that is, the fact that the complexity
$K(x,y)$ of the union of two strings $x$ and $y$ is not necessarily equal to the sum of
their individual complexities $K(x)$, $K(y)$. Rather, the joint complexity
satisfies the inequality
$K(x,y) \leq K(x) + K(y|x)$, where $K(x|y)$ is the length of
the shortest program that generates $y$ given $x$ as an input \cite{li1997}.
Applying this inequality recursively to the entire list shown in (1), it is not
surprising that the complexity of the list can be smaller than that of
some of its components.

In closing this comment, it may be of interest to point out that $\ell$ is equivalent
to the Champernowne constant~\cite{champ1933}, which is known to be random according
to the Shannon definition, that is, normal in the language of number theory,
and yet has a low algorithmic complexity \cite{li1997}.


\begin{thebibliography}{9}
\bibitem{abramo2005}  G. D'Abramo, Chaos, Solitons \& Fractals, 25 (2005) 29.
\bibitem{li1997}  M. Li, P. Vit\'{a}nyi, An Introduction to Kolmogorov
Complexity and Its Applications, New York: Springer, 1997.
\bibitem{champ1933} D.~G. Champernowne, J. London Math. Soc. 8 (1933) 254.
\end{thebibliography}
\end{document}